\documentclass[a4paper,12pt]{article}
\usepackage{psfig}
\usepackage{amssymb}

\def\bc{\begin{center}}
\def\ec{\end{center}}

\def\be{\begin{equation}}
\def\ee{\end{equation}}
\def\beq{\begin{eqnarray}}
\def\eeq{\end{eqnarray}}
\def\bfig{\begin{figure}}
\def\efig{\end{figure}}
\def\bnum{\begin{enumerate}}
\def\enum{\end{enumerate}}

\begin{document}
\begin{flushright}
{Journal-ref: Astronomy Reports, 2005, v.49, No.6, pp.470-476}
\end{flushright}
\bigskip
\bc
{\LARGE\bf Optimal Choice of the Softening Length and Time-Step in 
N-body Simulations}\\
\vspace{0.4cm}
{\bf S.A.~Rodionov\footnote{email:seger@astro.spbu.ru},
N.Ya.~Sotnikova}\\
\vspace{0.4cm}
{\it Sobolev Astronomical Institute, St. Petersburg State University,
Universitetskii pr.\\
28, Petrodvorets, St. Petersburg, 198904 Russia}\\

Received July 20, 2004; in final form, December 3, 2004 
\ec

\vspace{1cm}

A criterion for the choice of optimal softening length
$\epsilon$ and time-step $dt$ for
$N$-body simulations of a collisionless stellar system is analyzed.
Plummer and Hernquist spheres
are used as models to follow how changes in various parameters of an 
initially equilibrium stable model depend on $\epsilon$ and $dt$. 
These dependences are used to derive a criterion for choosing 
$\epsilon$ and $dt$. The resulting criterion is compared to Merritt's
criterion for choosing the softening length, which is based on
minimizing the mean irregular force acting on a particle with unit mass.
Our criterion for choosing $\epsilon$ and $dt$ indicate that $\epsilon$
must be a factor of 1.5-2 smaller than the mean distance between particles
in the densest regions to be resolved. The time-step must always
be adjusted to the chosen $\epsilon$ (the particle must, on average, 
travel a distance smaller than $0.5\epsilon$ during one
time-step). An algorithm for
solving N-body problems with adaptive variations of the softening length
is discussed in connection with the task of choosing $\epsilon$, but is
found not to be promising.\\

\medskip
\noindent
Key words: galaxies, numerical methods, N-body simulations


\section{Introduction}

The evolution of gravitating systems on time scale shorter than
two-body relaxation time is described by the
collisionless Boltzmann equation. The number of particles in $N$-body
simulations of such systems is usually several orders of magnitude
smaller than the number of stars in real systems. 
It could be said that in such N-body simulations collisionless Boltzman
equation is solved by Monte-Carlo method, in which the particles are used
to represent mass distribution of real system.
In practice,
integrating the equations of motion of the particles involves smoothing
the acutal (pointwise) potential of each individual particle, for example,
by substituting
a Plummer sphere potential for the actual potential. This procedure modifies
the law of gravity at small distances:
\beq
{\bf F}_{ij}^{\rm real} =
G m_i m_j \frac{{\bf r}_j - {\bf r}_i}{|{\bf r}_j - {\bf r}_i|^3}
~~~ \to ~~~
{\bf F}_{ij}^{\rm soft} = G m_i m_j \frac{{\bf r}_j - {\bf r}_i}
{(|{\bf r}_j - {\bf r}_i|^2 + \epsilon^2)^{3/2}} \, ,
\label{soft}
\eeq
where ${\bf F}_{ij}^{\rm real}$ and ${\bf F}_{ij}^{\rm soft}$ ---
are the real and softened forces,  respectively, acting on a particle
of mass 
$m_i$,  located at point  ${\bf r}_i$, produced by a particle of mass  $m_j$,
located at point ${\bf r}_j$, and $\epsilon$ --- is the softening length
determing the degree of potential smoothing.

Potential smoothing is used in $N$-body simulations for two reasons.

First, attempts to solve the equations of motion of gravitating points using
purely Newtonian potentials and simple constant-step integration always lead
to problems during close pair encounters (the integration scheme diverges).
Correct modeling of such systems requires the use of either variable-step
integrating algorithms or sophisticated regularization methods, which lead
to unreasonably long CPU times. No such problems arise when a softened
potential is used.

Second, smoothing the potential reduces the ``graininess'' of the particle
distribution, thereby making the potential of the model system more similar
to that of a system with a smooth density distribution, i.e., a system
described by the collisionless Boltzmann equation. 

It is obvious that $\epsilon$ cannot be too large: this would result in
substantial bias of the potential, and would also impose strong
constraints on the spatial resolution of structural features of the system.
It is important to have an objective criterion for choosing the softening
length in $N$-body simulations. In this paper, we derive such a criterion
by analyzing the time variations of the distribution functions for
spherically symmetric models with different values of $\epsilon$ .

\section{ Merritt's criterion for choosing $\epsilon$}

Merritt \cite{merrit-1996} proposed a criterion for choosing the softening
length based on the minimization of the mean irregular force acting on a
particle. He introduced the mean integrated square error ($MISE$),
which characterizes the difference between the force produced by a
discrete set of $N$ bodies and the force acting in a system with a continuous
density distribution:

\beq
MISE(\epsilon) &=& E \left( \int \rho ({\bf r})
|{\bf F}({\bf r},\epsilon)-{\bf F}_{\rm true}({\bf r})|^2 \,
d {\bf r} \right) \, ,
\eeq

Here, $E$ denotes averaging over many realizations of the system,
${\bf F}({\bf r},\epsilon)$  is the force acting on a particle with unit
mass located at the point ${\bf r}$  produced by $N$ particles with softened
potentials, and ${\bf F}_{\rm true}({\bf r})$  is the force acting on the
same particle in a system with a continuous density distribution $\rho(\bf r)$.

\begin{figure}
\centerline{\psfig{file=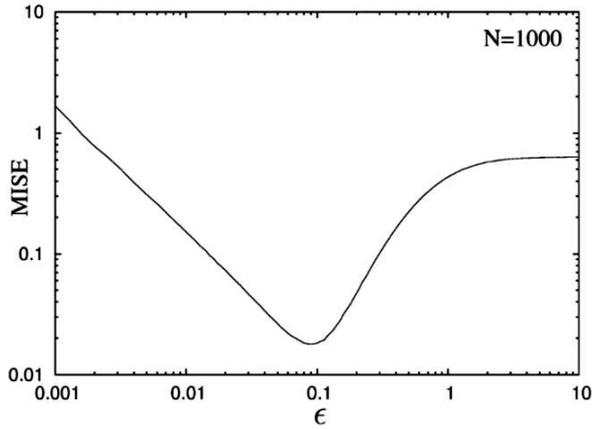,height=6cm}}
\caption[1]{Dependence of the $MISE$ on the softening length
$\epsilon$  for a Plummer model with $N = 1000$.}
\label{merritt-fig}
\end{figure}

\noindent
Two factors contribute to the $MISE$.
\begin{itemize}
\item[-]
Fluctuations of the discrete density distribution. These are very important at
small $\epsilon$, decrease with $\epsilon$ (see the left-hand branch of
the curve in Fig.~\ref{merritt-fig}),
and for a given $\epsilon$, decrease with increasing $N$ \cite{merrit-1996};
\item[-]
Errors in the representation of the potential (the difference between the
softened potential and a point-mass potential). These errors, on the contrary,
are very large for large $\epsilon$, decrease with decreasing $\epsilon$,
and do not depend on $N$ (the right-hand branch of the curve in 
Fig.~\ref{merritt-fig})
\end{itemize}

\noindent
The $MISE$ reaches its minimum at some $\epsilon = \epsilon_{m}$
(Fig.~\ref{merritt-fig}). Merritt suggested to adopt this $\epsilon_{m}$ as the
optimal choice of the softening length in $N$-body
simulations. This quantity depends on $N$ and, e.g., in
the case of a Plummer model, can be approximated
by the dependence $\epsilon_{m} \approx 0.58N^{-0.26}$ in the interval
$N$ from 30 to 300\,000 (in virial units, where $G=1$, the
total mass of the model is~$M_{\rm tot}=1$, and the total
energy of the system is~$E_{\rm tot}=-1/4$) \cite{ath-1998}.

\section{Formulation of the problem}
Merritt's criterion is based on minimizing the
mean error in the representation of the force in an
equilibrium system at the initial time. It is of interest
to derive a criterion for choosing the softening length
based directly on the dynamics of the system.
If a stable equilibrium configuration is described by
a discrete set of $N$ bodies, due to various errors, the
parameters of the system (e.g., the effective radius)
will deviate from their initial values with time. The
smaller these deviations, the more adequately the
model reproduces the dynamics of the system. It is
reasonable to suppose that these variations will be
minimized for some $\epsilon$.
We analyzed two equilibrium models: a Plummer sphere
model

\be
\rho(r) = \frac{3 M_{\rm tot}}{4\pi}\frac{a_{\rm P}^2}
{\left(r^2 + a_{\rm P}^2\right)^{5/2}}
\ee

\noindent
as an example of a model with a nearly homogeneous density
distribution, and a Hernquist-sphere model with an isotropic
velocity distribution \cite{hernq-1990}

\be
\rho(r) = \frac{M_{\rm tot}}{2\pi}\frac{a_{\rm H}}{r(r + a_{\rm H})^3}\, ,
\ee

\noindent
as an example of a model with more concentrated mass distribution.
In (3) and (4), $a_{\rm P}$  and $a_{\rm H}$ are
the scale lengths of the density distributions for the
Plummer and Hernquist models. (The Plummer and
Hernquist models contain about 35\% and 25\% of their
total masses inside the radii $a_{\rm P}$ and $a_{\rm H}$, respectively).
We use here virial units ($G=1,~M_{\rm tot}=1,~E_{\rm tot}=-1/4$),
for which $a_{\rm P} = 3\pi/16 \approx 0.59$ and $a_{\rm H} = 1/3$.
We computed the gravitation force using the $TREE$
method \cite{bh-1986}. In the computations, we set the parameter
$\theta$, which is responsible for the accuracy of the
computation of the force in the adopted algorithm,
equal to $0.75$. Test simulations made with higher ($\theta=0.5$) 
and lower ($\theta=1.0$) accuracy showed that our
conclusions are independent of $\theta$ within the indicated
interval. Hernquist et al. \cite{ hernq-1993} showed that the errors in
the representation of the interparticle force in the TREE method
are smaller than the errors due to discreteness noise, independent of $N$,
so that the above approach can be used
to solve the formulated problem. We integrated the
equations of motion of the particles using a standard
leap-frog scheme with second order accuracy in time-step. The
number of particles was $N = 10\,000$, and the computations
were carried out in the NEMO \cite{ teuben-1995} software
package.

We investigated two methods for smoothing the
potential. The first is based on substituting the Plummer
potential for the point-mass potential, and modifies the force of interaction between the particles
according to the scheme (\ref{soft}). The second method
uses a cubic spline \cite{ hernq-katz-1986} to smooth the potential. We
report here results only for the first method, but all our
conclusions remain unchanged in the case of spline
smoothing.

We analyzed how various global parameters of
the system are conserved, in particular, the variation
of the density distribution and, separately, the variation
of the particle-velocity distribution.

We characterized the deviation of the particle distribution
in space at time $t$  from the initial distribution
using the quantity $\Delta_r$ , which was computed follows.
We subdivided the model into spherical layers and
computed for each layer the difference between the
number of particles at time $t$ and the number at the
initial time. We then computed $\Delta_r$  as the sum of the
absolute values of these differences and normalized
it to $N$, the total number of particles in the system.
The thickness of the layers was $0.1$ and the maximum
radius was equal to $2$ (each of the spherical layers
in both models contained a statistically significant
number of particles, always greater than $50$). There is an important remark:
$\Delta_r$ will not be equal
to zero for two random realizations of the same model.
This quantity has appreciable natural noise due to the
finite number of particles considered. We estimated
the level of this natural noise to be the mean $\Delta_r$
averaged over a large number of pairs of random realizations
of the model.

We computed the parameter $\Delta_v$
for the distribution of particles in velocity space in a
similar way. For the results reported here, the thickness
of the layer was $0.1$ and the maximum velocity
was $1.5$ (this choice of maximum velocity ensured
that each spherical layer contained a statistically significant number of
particles for both models more than $30$).

We also computed the two-body relaxation time for the models as a function
of $\epsilon$ . We determined this times scale as the time
over which the particles deviate significantly from
their initial radial orbits (the deviation criterion and
method used to estimate it were similar to those
employed by Athanassoula et al. \cite{ath-2001}). The
estimation method can be briefly described as follows.
We constructed a random realization of the system
(the Plummer or Hernquist sphere). In this system,
we chose a particle moving toward the center in an
almost radial orbit and located about one effective
radius from the center. Then the orbit of this particle was touched up to 
strictly radial. We followed the evolution
of the entire system until this particle traveled a
distance equal to $1.5$ times its initial distance from
the center. Knowing the time $t_{\rm p}$ that has elapsed since
the start of the motion and the angle of deflection of
the particle from its initial radial orbit,
$\alpha_{\rm p}$, we can compute
the time required for the particle to deviate from
its initial direction by one radian. We then average
the resulting time over a large number of similar
particles to estimate the two-body relaxation time
\be
\frac{1}{t_{\rm relax}} =
\left<\frac{\alpha_{\rm p}}{\sqrt{t_{\rm p}}}\right>^2 \, ,
\label{relax}
\ee
where $t_{\rm relax}$ is the two-body relaxation time
and $<{\ldots}>$  denotes averaging over a large number
of test particles. Formula (\ref{relax}) is valid in the diffusion
approximation, i.e., assuming small angles for individual
scattering acts.

\section{ Results of numerical simulations}

\subsection{The Plummer Sphere}

\begin{figure}
\centerline{\psfig{file=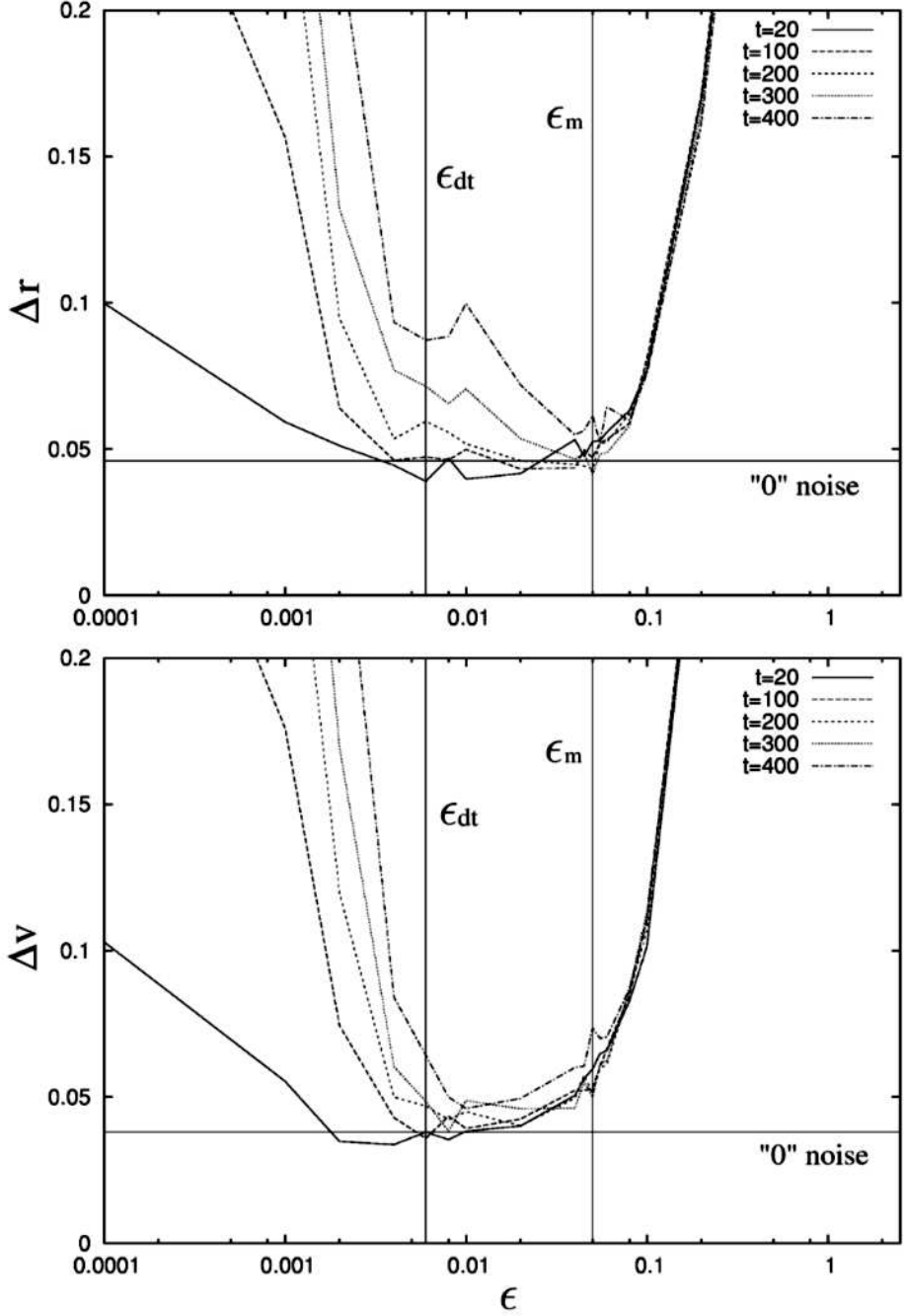,height=19cm}}
\caption[2]{ 
Dependence of $\Delta_r$  and $\Delta_v$ on $\epsilon$ for
different times for the Plummer-sphere model. The horizontal line shows
the natural noise levels for $\Delta_r$  and $\Delta_v$. The vertical
lines indicate the position of $\epsilon_m$ --- the optimal softening length
according to the criterion of Merritt 
(for $N = 10\,000$,  $\epsilon_m = 0.05$ in virial units), and the minimum
$\epsilon = \epsilon_{dt}$ to which the time-step was adjusted
($\epsilon_{dt} = 0.006$). Averaging was performed over several models
for each $\epsilon$. The time-step is $dt = 0.01$. }
\label{plum-fig}
\end{figure}

\begin{figure}
\centerline{\psfig{file=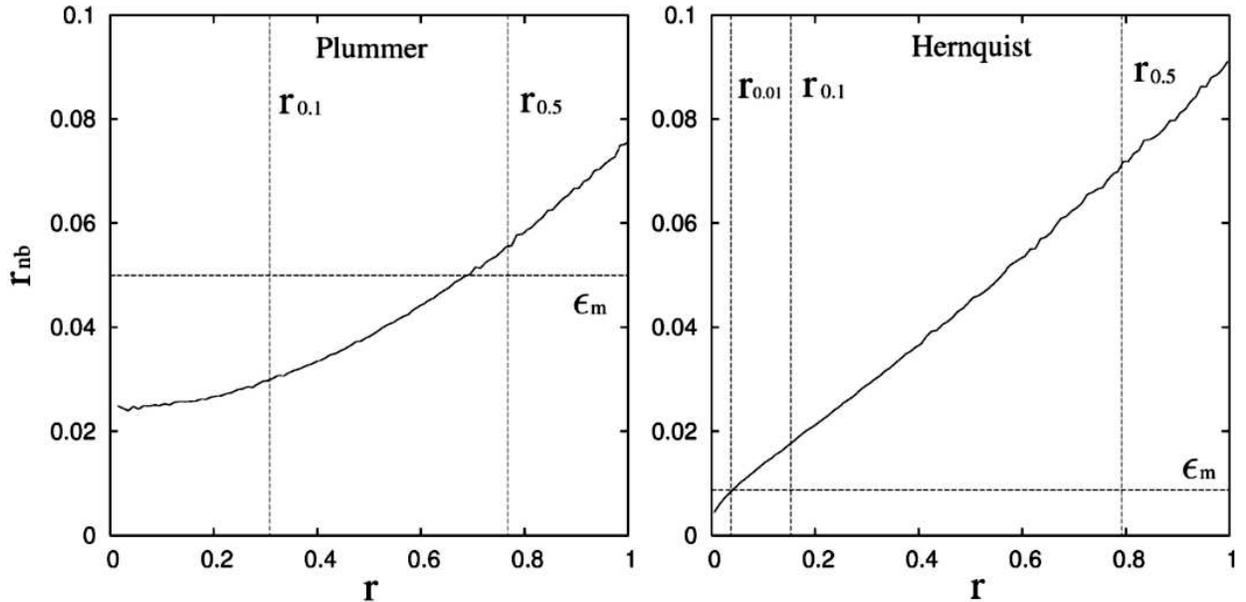,height=9cm}}
\caption[3]{ 
Average distance to the nearest neighbor at various radii for the Plummer
and Hernquist models with $N = 10\,000$. The vertical lines correspond to 
the radii of spheres containing 0.01, 0.1, and 0.5 of the total mass of the
system. The horizontal lines shows the optimal $\epsilon = \epsilon_m$
according to the criterion of Merritt. }
\label{fnb-fig}
\end{figure}

\begin{figure}
\centerline{\psfig{file=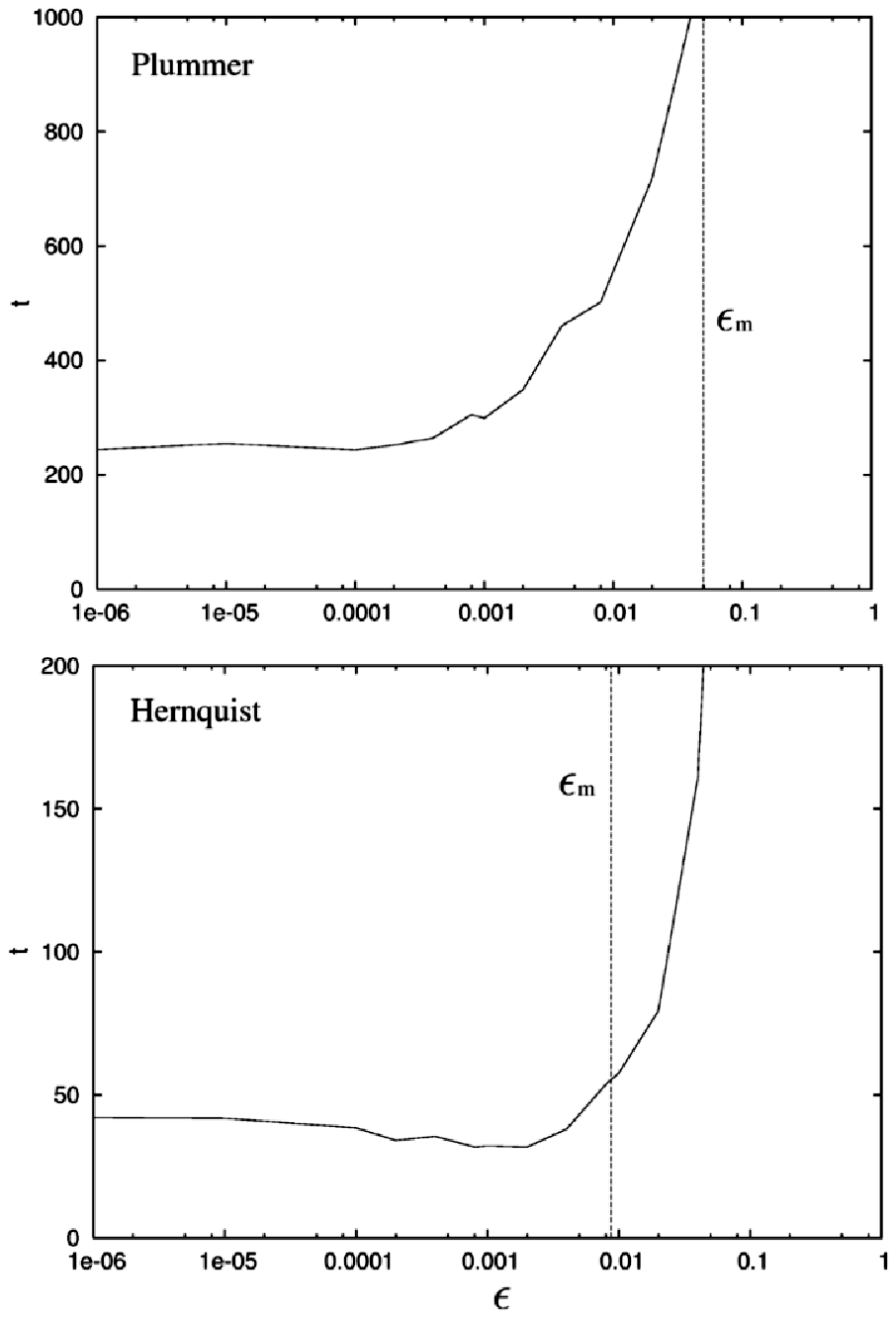,height=15cm}}
\caption[4]{
Relaxation times for the Plummer and Herquist
models as functions of $\epsilon$. The vertical line indicates the
optimum $\epsilon_m$ according to the criterion of Merritt. The time-step
was $dt = 0.001$. Averaging was performed
over 2500 and 1000 test particles for each $\epsilon$ for the Plummer
and Hernquist models, respectively.}
\label{trelax-fig}
\end{figure}

Figures \ref{plum-fig} - \ref{trelax-fig} illustrate the results of the
computations for the Plummer sphere. Our main conclusions
are the following.

\begin{enumerate}
\item
Deviations of the softened potential from the
Newtonian potential become important when $\epsilon > \epsilon_m$.
The system adjusts to the new potential by changing
its distribution function. The adjustment is almost
immediate, and occurs on time scales $t << 20$ (one
time unit corresponds approximately to the crossing
time for the core, $a_{\rm P}$). Such a strongly smoothed
system that has evolved far from its initial state shows
hardly any further changes (Fig.~\ref{plum-fig}). This result agrees
well with the behavior of the fractional error of the
total energy of the system for large $\epsilon$, discussed by
Athanassoula et al. \cite{ath-1998} (Fig.7 in their paper).

\item
Close pair encounters are computed incorrectly
when the particle travels a distance greater than $\epsilon$
during one time-step $dt$. This introduces an
error, which rapidly accumulates and results in substantial
changes in $\Delta_r$ and $\Delta_v$ (Fig.\ref{plum-fig}). This happens
when $\epsilon < \epsilon_{dt} = \sigma_v  dt$, where $\sigma_v$ is the
mean velocity dispersion in the system ($\sigma_v \approx 0.7$ 
in the adopted virial units for the Plummer model).

\item
The evolution of the system outside the interval
$\epsilon_{dt}<\epsilon<\epsilon_m$  is simulated incorrectly.
\item
A comparison of the relaxation time of the
system (Fig. \ref{trelax-fig}) with the time scale for the variations
of the density distribution in the system ($\Delta_r$) shows
that, in the interval  $\epsilon_{dt}<\epsilon<\epsilon_m$, $\Delta_r$
varies only insignificantly on time intervals a factor of two to three
shorter than the relaxation time. When $\epsilon = \epsilon_m$, the
system preserves its structure for the longest time,
since the relaxation time is maximum for this $\epsilon$ (upper
panel in Fig.~\ref{plum-fig}).
\item
At the same time, it is clear from Fig.~\ref{fnb-fig} that,
for the model considered, $\epsilon_m$ exceeds the mean distance
between particles in the central regions, so that
we expect strong modification of the potential in a
substantial fraction of the system at $\epsilon = \epsilon_m$. As they
adjust to the changed potential, the particles rapidly
change their dynamical characteristics ($\Delta_v$; lower
panel in Fig.~\ref{plum-fig}). It follows that $\epsilon_m$ is by no means
the best choice from the viewpoint of conserving the
initial velocity distribution function, although $\Delta_v$ remains
approximately constant after the adjustment
has ended.
\item
The effect of the two factors affecting the MISE
(the ``graininess'' of the potential and its modification
as a result of smoothing) are separated in time. The
first factor acts on time scales comparable to the
relaxation time (for a given $\epsilon$), whereas systematic
errors in the representation of the potential appear
almost immediately. It follows that if for the chosen $\epsilon$ 
systematic errors are small (i.e. $\epsilon$ is small enough) then the $N$-body
model very accurately represents the initial system, but only over a
time interval that is a factor of 2-3 shorter
than the relaxation time for the this $\epsilon$ 
(of course its right only if time-step was small enough, i.e.
$\epsilon > \epsilon_{dt}$).

\item
Choosing $\epsilon$ to be a factor of 1.5-2 smaller
than the mean distance between particles in the densest
region yields the optimal solution from the viewpoint
of conserving both $\Delta_r$  and $\Delta_v$. For the Plummer
sphere with $N = 10\,000$, this corresponds to $\epsilon \approx 0.01 - 0.02$ 
(in virial units). We can see from Fig.~\ref{plum-fig}
that the simulation of the system is almost optimal
when $\epsilon \approx 0.01-0.02$, but this is achieved by restricting
the evolution time of the system ( $t \leq 100$). When
$\epsilon=\epsilon_m \approx 0.05$, the system is simulated satisfactorily
on time scales  $t \leq 300$, but we must pay for this
increased evolution time: the system is not correctly
represented by the model from the very start.

\end{enumerate}

\subsection{The Hernquist Sphere}

Figures 3-5 illustrate the computation results for
the Hernquist-sphere model. Our main conclusions
are the following.

\begin{figure}
\centerline{\psfig{file=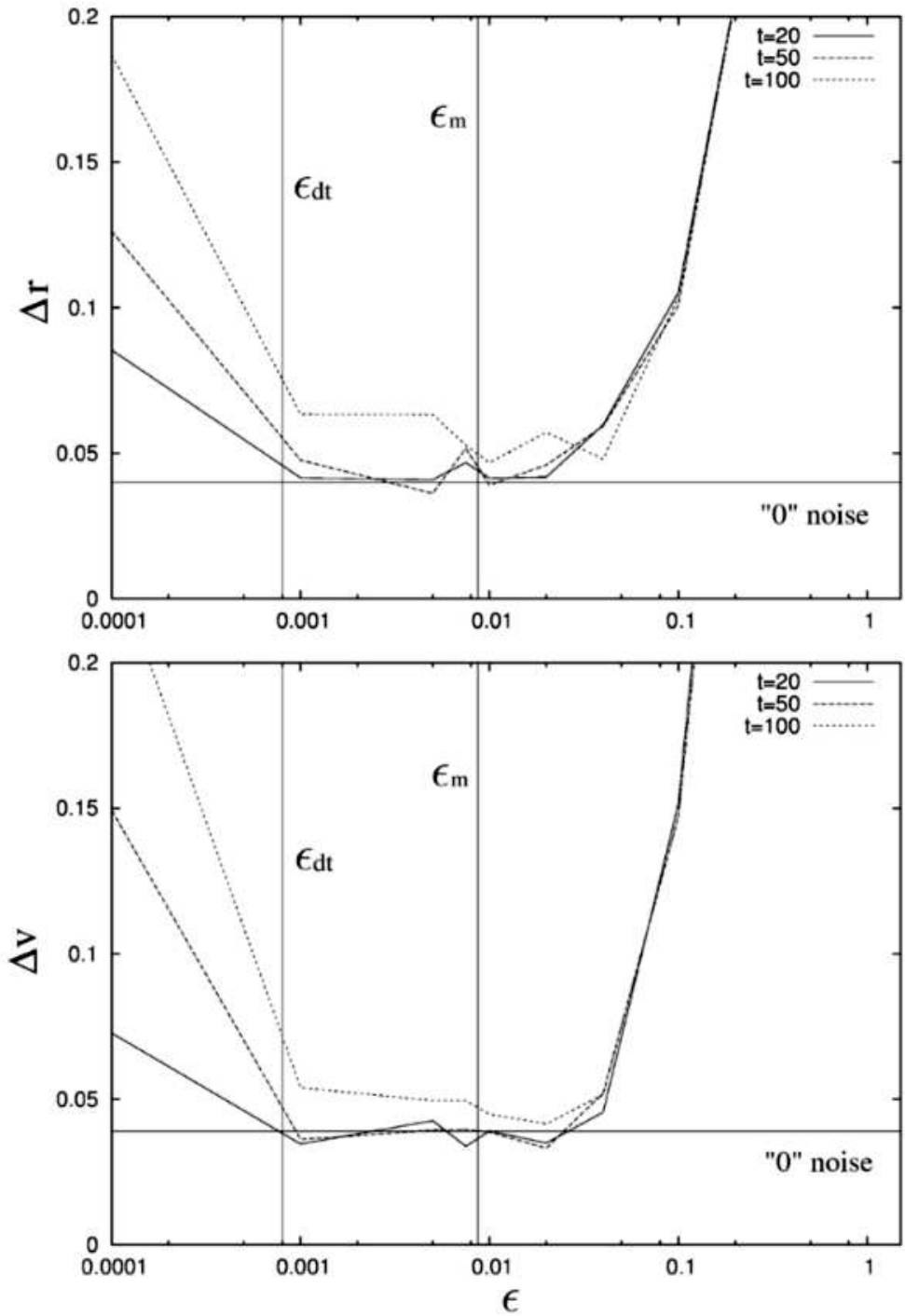,height=20cm}}
\caption[5]{
Same as Fig.~\ref{plum-fig} for the Hernquist-sphere. The time-step
is $dt = 0.001$. In this case, we have for $N = 10\,000$ and this
time-step $\epsilon_m = 0.0087$ $\epsilon_{dt} = 0.0008$. }
\label{hernq-fig}
\end{figure}

\begin{enumerate}
\item
We can see from Fig.~\ref{fnb-fig} that, with the
Hernquist-sphere model, $\epsilon_m$ is comparable to the
mean distance between particles in the very central
regions of the model (which contain less than 1\% of all
particles). This means that the MISE can be affected
even by a small number of particles with strongly
modified potentials.
\item
Figure \ref{hernq-fig} shows that the change in the particle
distributions in the configuration and velocity space
($\Delta_r$ and $\Delta_v$) are indistinguishable from the noise even
for $\epsilon>\epsilon_m \approx 0.0087$, right to $\epsilon \approx 0.02$.
This is explained by the fact that, when $\epsilon = 0.02$, 
only 10\% of the particles have $\epsilon$ values greater
than the mean distance between particles, and only this small number
of particles have strongly modified potentials. We simply do not observe
the adjustment of these central regions to the
new potential.

\item
We showed in the previous section that, for
nearly uniform models, $\epsilon$ should be chosen about a factor
of 1.5 - 2 smaller than the mean distance between
particles in the densest regions. This criterion cannot
be directly applied to significantly nonuniform models,
since the central density peaks would be washed
out due to the finite number of particles. For such
models, $\epsilon$ should be about a factor of 1.5 - 2 smaller than
the mean distance between particles in the regions
to be resolved. In the case of a Hernquist sphere
with $N = 10\,000$, the $N$-body model with $\epsilon = 0.01$
adequately represents the system everywhere except
for the most central regions, which contain 2-3\%
of all the particles (Fig.~\ref{fnb-fig}). When $\epsilon=0.02$,
already the
regions containing 10\% of all particles are not simulated
correctly.

\item
Unlike the Plummer sphere, in the case of the
Hernquist sphere, significant changes in $\Delta_r$ and $\Delta_v$
occur on times of the order of two ``relaxation times''
$t=100$ (Figs. 4 and 5). We put the term relaxation
time in quotes here, because this concept is not fully applicable
in the Hernquist model. The relaxation time
is, by definition, a local parameter, and has diferent
values in the central region and periphery for the
strongly nonuniform systems we analyze here.

\end{enumerate}

\section{Do N-body simulations need adaptive code?}

A number of authors have discussed the possibility
of developing a code for $N$-body simulations with
an adaptively adjustable softening length; i.e.,
with $\epsilon$ varying as a function of the mean distance
between particles at a given point of the system \cite{dehnen-2001},
\cite{merrit-1996}. Dehnen \cite{dehnen-2001} showed, based on the formalism
of Merritt \cite{merrit-1996}, that the use of such an adjustable 
softening length decreases significantly the role of the irregular
forces averaged over the entire system. It follows
from the results reported here that the only possible
advantage of such a code might be a reduction
of computing time when modeling very inhomogeneous
models. This follows from the fact that one can choose
larger softening lengths, and thereby longer time-steps,
in less dense regions.

We encountered a problem when implementing
such a code, however. The change of $\epsilon$ for individual
particle results in an asymmetric change of the total
energy of the system. When $\epsilon$ is increased 
($\epsilon_{\rm small} \to \epsilon_{\rm mean}$), particles located near
a radius of about $2\epsilon_{\rm mean}$ pass from a ``harder'' to a 
``softer'' potential, i.e., the absolute value of the potential energy
decreases. A decrease in $\epsilon$ 
($\epsilon_{\rm big} \to \epsilon_{\rm mean}$) results in the opposite
effect, but, in this case, on average, the change in the
potential influences the greater number of particles
located near the radius $2\epsilon_{\rm big}$, so that the absolute
value of the total potential energy of the system gradually
increases with time. Our $N$-body simulations
showed significant increases in the total energy. Various
artificial methods can be suggested to compensate
for this energy change, but it remains unclear
how this will affect the evolution of the system as a
whole.

\section{Conclusions}
The results of our study indicate that the softening length
$\epsilon$ in $N$-body simulations should be a
factor of $1.5-2$ smaller than the mean distance between
particles in the densest regions to be resolved.
In this case, the time-step must be adjusted to
the chosen $\epsilon$ (on average, the particle must travel a
distance smaller than one-half $\epsilon$ during one time-step).

The use of code with variable softening lengths
appears to be of limited value, and does not show
particular promise.\\

\vspace{0.5cm}
\noindent
{\large Acknowledgment}\\
This work was supported by the Russian Foundation
for Basic Research (project no. 03-12-17152), Federal Program ``Astronomy'' 
(grant no. 40.022.1.1.1101)
and by the President Russian Federation Program
for Support of Leading Scientific Schools of
Russia (grant no. NSh-1088.2003.2).

\begin{flushright}
Translated by A.Dambis
\end{flushright}


\end{document}